\documentclass{PoS}
\usepackage{overpic}
\usepackage{amsmath}
\usepackage{amstext}
\usepackage{relsize}
\usepackage{booktabs}
\usepackage{multirow}
\usepackage{xspace}

\input{babarsym}

\makeatletter
\DeclareRobustCommand\bfseries{%
  \not@math@alphabet\bfseries\mathbf
  \fontseries\bfdefault\selectfont\boldmath}

\makeatother

\newcommand{\superb}{Super\ensuremath{B}\xspace}
\renewcommand{\babar}{\mbox{%
    \slshape B\kern-0.1em{\smaller A}\kern-0.1em
    B\kern-0.1em{\smaller A\kern-0.2em R}}\xspace}

\newcommand{\EE}[1]{\ensuremath{\cdot 10^{#1}}}

\def\BR{{\ensuremath{\mathcal B}}\xspace}

\title{%
Search for Lepton-Flavor-Violating Tau Decays at the $B$-factories
}

\ShortTitle{%
Search for Lepton-Flavor-Violating Tau Decays at the $B$-factories
}

\author{\speaker{Alberto Lusiani}%
        \thanks{Representing the \babar collaboration}\\
       Scuola Normale Superiore and INFN -- Pisa\\
       E-mail: \email{alberto.lusiani@pi.infn.it}}

\abstract{%
Both the $B$-factories \babar and Belle have ended data-taking and have
mostly completed the data analysis aimed at searching Lepton Flavor
Violation in Tau decays. No evidence of LFV in tau decays has been
found yet. We review in the following the experimental upper limits
that have been set.}

\FullConference{The Xth Nicola Cabibbo International Conference on Heavy Quarks and Leptons,\\
		October 11-15, 2010\\
		Frascati (Rome) Italy}

\begin{document}

\section{Introduction}

The observation of neutrino oscillations implies that Lepton Flavor
Violation (LFV) must happen in the interactions and decays of the
charged leptons. Within the Standard Model (SM), however, LFV decays
of charged leptons are heavily suppressed with respect to the dominant
processes by the fourth power of the small ratio between the neutrino mass
differences and the $W$ mass, $\Delta m_\nu^4/m_W^4$. On the other
hand, most models beyond the SM, such as
Supersymmetry, Grand Unification or Extra Dimensions, naturally include
additional lepton-flavor-violating processes, which can produce
observable tau decay rates with LFV, even when the model parameters
are constrained with the available experimental
data~\cite{Ellis:1999uq,Ellis:2002fe,Masiero:2002jn,Fukuyama:2003hn,Cvetic:2002jy,Yue:2002ja,Dedes:2002rh,Brignole:2003iv,Arganda:2005ji}.

In the past, the CLEO collaboration has searched for LFV in tau decays
without finding any evidence, and has set 90\% CL upper limits around
$10^{-6}$ for the corresponding tau branching fractions. In the last
decade, the \babar and Belle $B$-factories collaborations have extensively
searched for LFV in tau decays, exploiting two orders of magnitude
larger samples of \epem annihilation data at and around the \FourS
peak, with a sensitivity for tau LFV branching fractions up to about
$10^{-8}$.

The \babar collaboration has ended data-taking in April 2008,
collecting about 531\invfb of data, and has completed its data
analysis on about its full data sample for the most interesting LFV
channels, $\tau\to (\mu/e) \gamma$ and $\tau\to 3\ell$, and on samples
close to it full data samples for the other channels.

The Belle collaboration has ended data taking in June 2010, collecting
about 1040\invfb of data, has recently produced results on some tau
LFV searches using most of its complete dataset, and is updating to
its final dataset the most complex searches, such as $\tau\to (\mu/e)
\gamma$.

We report in the following on the most recent $B$-factories results,
and we refer the reader to the HFAG group 2010
report~\cite{TheHeavyFlavorAveragingGroup:2010qj} for a comprehensive
status of tau LFV searches.

The typical search for LFV tau decays selects candidate $\tau^+\tau^-$
events where one tau decays into the most common final states, while
the other one decays into a neutrinoless final state with a specific
particle content. With respect to \qqbar events, $\tau^+\tau^-$ events
have fewer tracks and are more collimated around the thrust axis in
the \epem center-of-mass (CM) reference frame.
Candidate events from di-lepton final states are recognized because
their total energy is close to the \epem energy, while for
candidate $\tau^+\tau^-$ pairs a significant share of the energy escapes
undetected with one or two neutrinos of the non-LFV tau
decay. Finally, two-photon events can be recognized because their
total energy is smaller than for $\tau^+\tau^-$ events, with zero net
transverse momentum and with a larger imbalance of momentum along the
beams in the \epem CM system.

While standing in the CM system, both tau decay products cluster in
two relatively collimated cones around the thrust axis, aligned with
their respective tau directions. Since the LFV tau decay is
neutrinoless, the particles in its hemisphere have an invariant mass
that matches the tau mass within the experimental resolution and a
total energy that matches half the \epem energy ($\sqrt{s}\approx
10.58\gev$).  The experimental resolution in the invariant mass can be
improved by constraining the total energy to half the event energy
and by recovering Bremsstrahlung radiation that is close enough to the
final state tracks, obtaining a resolution of $10{-}20\mev$.  The
experimental resolution in the final state energy is around $40\mev$.

Selected candidates whose energy and invariant mass match the
expected values are counted to detect evidence for a signal exceeding
the expected background. The background is estimated by counting for
each background source how many Monte Carlo simulated events survive
the selection procedure. For some background sources, such as Bhabha and
two-photon events, it is too resource-demanding or otherwise
inconvenient to produce large enough simulated samples: in these cases
properly selected data control samples are used to estimate the
background. The selection procedure is optimized without looking at
data where the LFV signal is expected, to avoid the experimenter bias;
\babar usually optimizes the selection to obtain the lowest expected
90\% CL upper limit in case there is no signal, while Belle usually
optimizes in order to be able to detect the smallest possible tau LFV
branching fraction with 99\% CL evidence.

\section{Search for the tau decay into a lepton and neutral
  pseudoscalar meson}

Belle has presented preliminary results on a search for $\tau\to\ell
P^0$, $P^0{=}\pi^0,\eta,\eta^\prime$~\cite{Hayasaka:2010et}, which
updates a 2007 publication based on about
400\invfb~\cite{Miyazaki:2007jp} (\babar also published results on a
similar sample~\cite{Aubert:2006cz}).  All searches did not find any
evidence of a LFV signal.

In the most recent Belle analysis, the eta mesons have been searched
on two decay modes each, $\eta\to \gamma\gamma,\; \pi^+\pi^-\pi^0$ and
$\eta^\prime\to \pi^+\pi^-\eta,\; \rho^0\gamma$.  The selection has been
further optimized and reports a signal efficiency 1.5 higher with less
than one expected background event for each mode.
The 90\% CL upper limits on the branching fraction are
$(2.2-4.4){\times}10^{-8}$ and are detailed in Table.~\ref{table:lp0}.

\begin{table}[tb]
\begin{center}\smaller
\begin{tabular}{lccclccc}
\toprule
 Mode ($\tau\rightarrow$) & Eff.(\%) &$N_{\text{BG}}^{\text{exp}}$ & UL
 (${{\times}10}^{-8}$)& 
 Mode ($\tau\rightarrow$) & Eff.(\%) &$N_{\text{BG}}^{\text{exp}}$ & UL
 (${{\times}10}^{-8}$)\\
\midrule
$\mu\eta(\rightarrow\gamma\gamma)$ &8.2&$0.63\pm0.37$&3.6&
$e\eta(\rightarrow\gamma\gamma)$ &7.0&$0.66\pm0.38$&8.2\\
$\mu\eta(\rightarrow\pi\pi\pi^0)$ &6.9&$0.23\pm0.23$&8.6&
$e\eta(\rightarrow\pi\pi\pi^0)$ &6.3&$0.69\pm0.40$&8.1\\
\midrule
$\mu\eta$(comb.) &&&2.3&
$e\eta$(comb.) &&&4.4\\
\midrule
$\mu\eta'(\rightarrow\pi\pi\eta)$ &8.1&$0.00^{+0.16}_{-0.00}$
&10.0&
$e\eta'(\rightarrow\pi\pi\eta)$ &7.3&$0.63\pm0.45$&9.4\\
$\mu\eta'(\rightarrow\gamma\rho^0)$ &6.2&$0.59\pm0.41$&6.6&
$e\eta'(\rightarrow\gamma\rho^0)$ &7.5&$0.29\pm0.29$&6.8\\
\midrule
$\mu\eta'$(comb.) &&&3.8&
$e\eta'$(comb.) &&&3.6\\
\midrule
$\mu\pi^0$ &4.2&$0.64\pm0.32$&2.7&
$e\pi^0$ &4.7&$0.89\pm0.40$&2.2\\
\bottomrule
\end{tabular}
\end{center}
\vspace{-0.5\baselineskip}
\caption{Search for $\tau\to\ell P^0$, $P^0{=}\pi^0,\eta,\eta^\prime$,
  efficiencies (Eff.), expected number of background events
  ($N_{\text{BG}}^{\text{exp}}$), 90\% CL upper limits on the branching fractions
  (UL), where (comb.) indicated the combined upper limit when
  combining all examined modes of the eta mesons decay.}
\label{table:lp0}
\end{table}

\section{Search for the tau decay into a lepton and neutral
  vector meson}

Belle has presented preliminary results on searches for $\tau\to\ell
V^0$, $V^0{=}\rho,K^{*0},\omega,\phi$ on a sample of 854\invfb of
data~\cite{Hayasaka:ichep-2010}, updating on former published results
based on 543\invfb~\cite{Nishio:2008zx}. No evidence for a signal has
been found, as well as in the \babar searches on the same channels
that were based on 451\invfb of
data~\cite{Aubert:2007kx,Aubert:2009my}.

The selection has been improved, and on average the signal efficiency
has been increased by 20\%, while maintaining the expected background
at around one event or less for all channels. Backgrounds have been
studied in higher detail and it was found that two-photon and
radiative Bhabha processes constitute a non-negligible background for
$\tau\to\mu\rho$. Table~\ref{table:lv0} reports the established 90\%
CL upper limits.

\begin{table}[tb]
  \begin{center}\smaller
    \begin{tabular}{lccclccc}
      \toprule
      Mode ($\tau\rightarrow$) & Eff.(\%) &$N_{\text{BG}}^{\text{exp}}$ & UL (${\times}10^{-8}$) & 
      Mode ($\tau\rightarrow$) & Eff.(\%) &$N_{\text{BG}}^{\text{exp}}$ & UL (${\times}10^{-8}$) \\
      \midrule
      $e\rho$   & 7.6\% & $0.29\pm0.15$ & 1.8 & $e K^{*0}$              & 4.4\% & $0.39\pm0.14$ & 3.2 \\
      $\mu\rho$ & 7.1\% & $1.48\pm0.35$ & 1.2 & $\mu K^{*0}$            & 3.4\% & $0.53\pm0.20$ & 7.2 \\
      $e\phi$   & 4.2\% & $0.47\pm0.19$ & 3.1 & $e \overline{K^{*0}}$   & 4.4\% & $0.08\pm0.08$ & 3.4 \\
      $\mu\phi$ & 3.2\% & $0.06\pm0.06$ & 8.4 & $\mu \overline{K^{*0}}$ & 3.6\% & $0.45\pm0.17$ & 7.0 \\
      $e\omega$ & 2.9\% & $0.30\pm0.14$ & 4.8 & $\mu\omega$             & 2.4\% & $0.72\pm0.18$ & 4.7 \\
      \bottomrule
    \end{tabular}
  \end{center}
  \vspace{-0.5\baselineskip}
  \caption{Search for $\tau\to\ell
    V^0$, $V^0{=}\rho,K^{*0},\omega,\phi$, efficiencies (Eff.),
    expected numbers of background events ($N_{\text{BG}}^{\text{exp}}$), 
    90\% CL upper limit on the branching fraction (UL) for each mode.}
  \label{table:lv0}
\end{table}

\section{Search for the tau decay into three leptons}

Both \babar and Belle have published in 2010 results on searches for $\tau\to
3\ell$~\cite{Lees:2010ez,Hayasaka:2010np}.
\babar has examined 486\invfb of data, most of its dataset, and has
considerably improved the previously published
result~\cite{Aubert:2003pc}, based on 221\invfb, by using improved
reconstruction software, especially regarding particle identification,
where for instance the muon efficiency has increased from 66\% to
77\%, and by more carefully optimizing the event selection.
Belle has updated a former search based on 535\invfb of
data~\cite{Miyazaki:2007zw} to 782\invfb, maintaining a
selection with a similar efficiency and high background suppression,
with at most 0.2 events expected in all channels.
Both experiments found no evidence for LFV in all channels,
Table~\ref{table:3l} lists the 90\% CL upper limits.

\begin{table}[h]
\begin{center}\smaller
\begin{tabular}{lcccccc}
\toprule
& \multicolumn{3}{c}{Belle} & \multicolumn{3}{c}{\babar} \\
Mode ($\tau^-\rightarrow$) & Eff.(\%) &$N_{\text{BG}}^{\text{exp}}$ & UL (${\times}10^{-8}$)
                           & Eff.(\%) &$N_{\text{BG}}^{\text{exp}}$ & UL (${\times}10^{-8}$)\\
\midrule
$e^-e^+e^-$   &6.0&$0.21\pm0.15$&2.7&  8.6&$0.12\pm0.02$&3.4 \\
$e^-e^+\mu^-$ &9.3&$0.04\pm0.04$&1.8&  8.8&$0.64\pm0.19$&3.7 \\
$e^-\mu^+e^-$ &11.5&$0.01\pm0.01$&1.5& 12.6&$0.34\pm0.12$&2.2 \\
$e^-\mu^+\mu^-$ &6.1&$0.10\pm0.04$&2.7& 6.4&$0.54\pm0.14$&4.6 \\
$\mu^-e^+\mu^-$ &10.1&$0.02\pm0.02$&1.7& 10.2&$0.03\pm0.02$&2.8 \\
$\mu^-\mu^+\mu^-$&7.6&$0.13\pm0.06$&2.1& 6.6&$0.44\pm0.17$&4.0 \\
\bottomrule
\end{tabular}
\end{center}
\vspace{-0.5\baselineskip}
\caption{Search for $\tau\to3\ell$, Belle and \babar efficiencies (Eff.),
expected numbers of background events ($N_{\text{BG}}^{\text{exp}}$), 
90\% CL upper limit on the branching fraction (UL) for each mode.}
\label{table:3l}
\end{table}

\section{Future prospects}

The $B$-factories \babar and Belle have both ended data-taking and are
close to complete tau LFV searches with a sensitivity for branching
fractions up to about $10^{-8}$, a two orders of magnitude improvement
with respect to the previously existing limits set by CLEO.  There are
good prospects that within the next ten years the proposed super
flavor factories BelleII~\cite{Aushev:2010bq} and
\superb~\cite{Bona:2007qt} will collect about 100 times larger samples
of \epem annihilations around the \FourS peak, permitting a
substantial advancement of the experimental sensitivity to the tau
LFV. If the machine backgrounds can be kept under control, as it
appears possible especially if the nano-beams design option is
followed, just repeating the $B$ factories analyses unchanged will
provide 90\% CL upper limits that scale as the inverse square root of
the integrated luminosity (${\propto}1/\sqrt{\cal L}$). By optimizing
the event selection for the larger datasets, however, the sensitivity
to tau LFV will improve at a faster pace, up to ${\propto}1/{\cal
L}$ for those searches that are background-free, i.e.\ whose candidate
selection can be optimized to retain the $B$-factories efficiencies
while at the same time suppressing backgrounds to expect at most one
background event in the signal region.
The \superb collaboration has recently produced a physics
report~\cite{Abe:2010sj} where the sensitivities for the search for
LFV in $\tau\to (e/\mu)\gamma$ and $\tau\to 3\ell$ are estimated.

The last published \babar results on
$\tau\to(\mu/e)\gamma$~\cite{Aubert:2009tk} are extrapolated to the
\superb expected luminosity assuming that the analysis is
background-dominated, i.e.\ that the background cannot be further
suppressed at constant signal efficiency. Improvements in the \superb
detector project with respect to \babar will however provide a better
sensitivity than the ${\propto}1/\sqrt{\cal L}$ extrapolation.  The
\superb detector will have better tracking resolution, to compensate
for the smaller boost, and its beam spot will be smaller: both
features improve the resolution on the energy-constrained invariant
mass and on the energy. Improvements are also expected in the photon
acceptance. The \superb sensitivity for $\tau\to\mu\gamma$ is
determined to be $2.4\EE{-9}$ as 90\% CL upper limit and $5.4\EE{-9}$
for a $3\sigma$ signal evidence. For the $\tau\to e\gamma$ the two
figures are $3.0\EE{-9}$ and $6.8\EE{-9}$, respectively.

The last published \babar analysis on $\tau\to
3\ell$~\cite{Lees:2010ez} is re-optimized for the \superb design
integrated luminosity of 75\invab, disregarding further gains from
detector improvements, and the \superb LFV sensitivity is determined
to be in the range $2.3{-}8.2\EE{-10}$ as 90\% CL upper limits and
$1.2{-}4.0\EE{-9}$  for a $3\sigma$ signal evidence.

\bibliographystyle{JHEP}
\bibliography{%
  belle-2006-2008,
  biblio-superb,
  ichep10-bib,
  hql10-bib,
  pub-a-pre90,
  pub-b-90-94,
  pub-c-95-99,
  pub-d-00-04,
  pub-e-05-09,
  pub-extra,
  tau-lepton
}

\providecommand{\href}[2]{#2}\begingroup\raggedright\begin{thebibliography}{10}

\bibitem{Ellis:1999uq}
J.~R. Ellis, M.~E. Gomez, G.~K. Leontaris, S.~Lola, and D.~V. Nanopoulos, {\it
  {Charged lepton flavour violation in the light of the Super-Kamiokande
  data}},  {\em Eur. Phys. J.} {\bf C14} (2000) 319--334,
  [\href{http://xxx.lanl.gov/abs/hep-ph/9911459}{{\tt hep-ph/9911459}}].

\bibitem{Ellis:2002fe}
J.~R. Ellis, J.~Hisano, M.~Raidal, and Y.~Shimizu, {\it {A new parametrization
  of the seesaw mechanism and applications in supersymmetric models}},  {\em
  Phys. Rev.} {\bf D66} (2002) 115013,
  [\href{http://xxx.lanl.gov/abs/hep-ph/0206110}{{\tt hep-ph/0206110}}].

\bibitem{Masiero:2002jn}
A.~Masiero, S.~K. Vempati, and O.~Vives, {\it {Seesaw and lepton flavour
  violation in SUSY SO(10)}},  {\em Nucl. Phys.} {\bf B649} (2003) 189--204,
  [\href{http://xxx.lanl.gov/abs/hep-ph/0209303}{{\tt hep-ph/0209303}}].

\bibitem{Fukuyama:2003hn}
T.~Fukuyama, T.~Kikuchi, and N.~Okada, {\it {Lepton flavor violating processes
  and muon g-2 in minimal supersymmetric SO(10) model}},  {\em Phys. Rev.} {\bf
  D68} (2003) 033012, [\href{http://xxx.lanl.gov/abs/hep-ph/0304190}{{\tt
  hep-ph/0304190}}].

\bibitem{Cvetic:2002jy}
G.~Cvetic, C.~Dib, C.~S. Kim, and J.~D. Kim, {\it {On lepton flavor violation
  in tau decays}},  {\em Phys. Rev.} {\bf D66} (2002) 034008,
  [\href{http://xxx.lanl.gov/abs/hep-ph/0202212}{{\tt hep-ph/0202212}}].

\bibitem{Yue:2002ja}
C.-x. Yue, Y.-m. Zhang, and L.-j. Liu, {\it {Non-universal gauge bosons Z' and
  lepton flavor-violation tau decays}},  {\em Phys. Lett.} {\bf B547} (2002)
  252--256, [\href{http://xxx.lanl.gov/abs/hep-ph/0209291}{{\tt
  hep-ph/0209291}}].

\bibitem{Dedes:2002rh}
A.~Dedes, J.~R. Ellis, and M.~Raidal, {\it {Higgs mediated B/(s,d)0 --> mu tau,
  e tau and tau --> 3mu, e mu mu decays in supersymmetric seesaw models}},
  {\em Phys. Lett.} {\bf B549} (2002) 159--169,
  [\href{http://xxx.lanl.gov/abs/hep-ph/0209207}{{\tt hep-ph/0209207}}].

\bibitem{Brignole:2003iv}
A.~Brignole and A.~Rossi, {\it {Lepton flavour violating decays of
  supersymmetric Higgs bosons}},  {\em Phys. Lett.} {\bf B566} (2003) 217--225,
  [\href{http://xxx.lanl.gov/abs/hep-ph/0304081}{{\tt hep-ph/0304081}}].

\bibitem{Arganda:2005ji}
E.~Arganda and M.~J. Herrero, {\it {Testing supersymmetry with lepton flavor
  violating tau and mu decays}},  {\em Phys. Rev.} {\bf D73} (2006) 055003,
  [\href{http://xxx.lanl.gov/abs/hep-ph/0510405}{{\tt hep-ph/0510405}}].

\bibitem{TheHeavyFlavorAveragingGroup:2010qj}
{The Heavy Flavor Averaging Group}, {\it {Averages of b-hadron, c-hadron, and
  tau-lepton Properties}},  \href{http://xxx.lanl.gov/abs/1010.1589}{{\tt
  arXiv:1010.1589}}.

\bibitem{Hayasaka:2010et}
K.~Hayasaka, {\it {Recent Tau Decay Results at B Factories -- Lepton Flavor
  Violating Tau Decays}},  \href{http://xxx.lanl.gov/abs/1010.3746}{{\tt
  arXiv:1010.3746}}.

\bibitem{Miyazaki:2007jp}
{\bf {Belle}} Collaboration, Y.~Miyazaki {\em et.~al.}, {\it {Search for lepton
  flavor violating tau- decays into l- eta, l- eta' and l- pi0}},  {\em Phys.
  Lett.} {\bf B648} (2007) 341--350,
  [\href{http://xxx.lanl.gov/abs/hep-ex/0703009}{{\tt hep-ex/0703009}}].

\bibitem{Aubert:2006cz}
{\bf {BaBar}} Collaboration, B.~Aubert {\em et.~al.}, {\it {Search for Lepton
  Flavor Violating Decays $\tau^\pm \to \ell^\pm \pi^0$, $\ell^\pm \eta$,
  $\ell^\pm \eta^\prime$}},  {\em Phys. Rev. Lett.} {\bf 98} (2007) 061803,
  [\href{http://xxx.lanl.gov/abs/hep-ex/0610067}{{\tt hep-ex/0610067}}].

\bibitem{Hayasaka:ichep-2010}
K.~Hayasaka, ``{Search for Lepton Flavour Violating tau decay and lepton-number
  violation B decay at Belle}.'' Talk given at the 35th International
  Conference on High Energy Physics, Paris, July, 2010.

\bibitem{Nishio:2008zx}
{\bf {Belle}} Collaboration, Y.~Nishio {\em et.~al.}, {\it {Search for
  lepton-flavor-violating $\tau\to\ell V^0$ decays at Belle}},  {\em Phys.
  Lett.} {\bf B664} (2008) 35--40,
  [\href{http://xxx.lanl.gov/abs/0801.2475}{{\tt arXiv:0801.2475}}].

\bibitem{Aubert:2007kx}
{\bf {BaBar}} Collaboration, B.~Aubert {\em et.~al.}, {\it {Search for Lepton
  Flavor Violating Decays $\tau^\pm \to \ell^\pm\omega$ ($\ell = e, \mu$)}},
  {\em Phys. Rev. Lett.} {\bf 100} (2008) 071802,
  [\href{http://xxx.lanl.gov/abs/0711.0980}{{\tt arXiv:0711.0980}}].

\bibitem{Aubert:2009my}
{\bf {BaBar}} Collaboration, B.~Aubert {\em et.~al.}, {\it {Improved limits on
  lepton flavor violating tau decays to l phi, l rho, l K* and l K*bar}},  {\em
  Phys. Rev. Lett.} {\bf 103} (2009) 021801,
  [\href{http://xxx.lanl.gov/abs/0904.0339}{{\tt arXiv:0904.0339}}].

\bibitem{Lees:2010ez}
{\bf {BaBar}} Collaboration, J.~P. Lees {\em et.~al.}, {\it {Limits on tau
  Lepton-Flavor Violating Decays in three charged leptons}},  {\em Phys. Rev.}
  {\bf D81} (2010) 111101, [\href{http://xxx.lanl.gov/abs/1002.4550}{{\tt
  arXiv:1002.4550}}].

\bibitem{Hayasaka:2010np}
K.~Hayasaka {\em et.~al.}, {\it {Search for Lepton Flavor Violating Tau Decays
  into Three Leptons with 719 Million Produced Tau+Tau- Pairs}},  {\em Phys.
  Lett.} {\bf B687} (2010) 139--143,
  [\href{http://xxx.lanl.gov/abs/1001.3221}{{\tt arXiv:1001.3221}}].

\bibitem{Aubert:2003pc}
{\bf BaBar} Collaboration, B.~Aubert {\em et.~al.}, {\it {Search for lepton
  flavor violation in the decay $\tau^- \to \ell^- \ell^+ \ell^-$}},  {\em
  Phys. Rev. Lett.} {\bf 92} (2004) 121801,
  [\href{http://xxx.lanl.gov/abs/hep-ex/0312027}{{\tt hep-ex/0312027}}].

\bibitem{Miyazaki:2007zw}
{\bf {Belle}} Collaboration, Y.~Miyazaki {\em et.~al.}, {\it {Search for Lepton
  Flavor Violating tau Decays into Three Leptons}},  {\em Phys. Lett.} {\bf
  B660} (2008) 154--160, [\href{http://xxx.lanl.gov/abs/0711.2189}{{\tt
  arXiv:0711.2189}}].

\bibitem{Aushev:2010bq}
T.~Aushev {\em et.~al.}, {\it {Physics at Super B Factory}},
  \href{http://xxx.lanl.gov/abs/1002.5012}{{\tt arXiv:1002.5012}}.

\bibitem{Bona:2007qt}
M.~Bona {\em et.~al.}, {\it {SuperB: A High-Luminosity Asymmetric e+ e- Super
  Flavor Factory. Conceptual Design Report}},
  \href{http://xxx.lanl.gov/abs/0709.0451}{{\tt arXiv:0709.0451}}.

\bibitem{Abe:2010sj}
T.~Abe {\em et.~al.}, {\it {Belle II Technical Design Report}},
  \href{http://xxx.lanl.gov/abs/1011.0352}{{\tt arXiv:1011.0352}}.

\bibitem{Aubert:2009tk}
{\bf {BaBar}} Collaboration, B.~Aubert {\em et.~al.}, {\it {Searches for Lepton
  Flavor Violation in the Decays tau -> e gamma and tau -> mu gamma}},  {\em
  Phys. Rev. Lett.} {\bf 104} (2010) 021802,
  [\href{http://xxx.lanl.gov/abs/0908.2381}{{\tt arXiv:0908.2381}}].

\end{thebibliography}\endgroup

\end{document}